\magnification=\magstep1
 \input vanilla.sty 
 \input pictex
 \font\tiny=cmr7
 =
    cmbx12   \font\small=cmr9
\font\smit=cmti9
\font\smsl=cmsl9    
\font\smmit=cmmi9
\font\smbf=cmbx9
\font\ninesy=cmsy9

\def\smalltype{\baselineskip=10pt \small
\textfont1=\smmit \textfont2=\ninesy 
\textfont\rmfam=\small
\textfont\bffam=\smbf \textfont\itfam=\smit \textfont\slfam=\smsl
\setbox\strutbox=\hbox{\vrule height7.5pt depth2.5pt width0pt}
\def\rm{\fam0\small}
\def\it{\fam\itfam\smit}
\def\sl{\fam\slfam\smsl}
\def\bf{\fam\bffam\smbf}
}

\def\mod{\mathop{\%}}
\def\<{\noindent }
 \def\zB{\cdot}  
 \def\ze{{{}\ifmmode\epsilon\else$\epsilon$ \fi}}
 \def\zH{\equiv} 
 \def\zj{\quad}   
 \def\zJ{\qquad}    
 \def\zM{\times} 
 \def\zp{{{}\ifmmode\pi\else$\pi$ \fi}}
 \def\zT#1{{\text{#1}}}    
 \def\zy{{{}\ifmmode\theta\else$\theta$ \fi}}

 \def\z#1#2{\ifcase#1 {\overline {#2}} \or                   
 {\null\ifmmode{\underline #2}\else{\underbar #2}\fi} \or    
 {\if #2i {\hat\imath } \else\if #2j {\hat\jmath }           
        \else {\hat #2} \fi\fi} \or
 {\if #2i {\vec\imath} \else\if #2j {\vec\jmath}             
        \else {\vec #2} \fi\fi} \or
 {\if #2i {\tilde\imath} \else\if #2j {\tilde\jmath}         
        \else {\tilde #2} \fi\fi} \or
 {{\text{\b{$#2$}}}} \or   
 {\!^{#2}} \or  
 {\!_{#2}} \or  
 {{\text{*}}#2} \or        
 {{\bold #2}} \fi}                                           



{\advance\baselineskip1\jot
\title\medbold
   Large Numbers, the 
Chinese Remainder Theorem,\\
 \medbold and the Circle of Fifths
\endtitle}

  \rightline{S. A. FULLING}
 \rightline{\small Texas A\&M University}
 \rightline{\small College Station, TX 77843-3368}
 \bigskip

{\<\bf The problem\quad}
The sort of problem that got me interested in this subject 
 is something like this:
Consider a recurrence relation such as
$$ a_{n+1} = a_n\z62 + (n+3)n a_n\,, \zJ a_0 = 1,$$
whose solutions are integers that grow rapidly with~$n$.
 (This is a cooked-up example.
 For an indication of a realistic problem, look at [2], especially 
the formula at the end, and contemplate calculating $Y_{40}\,$.)
Suppose that:
(1) We do not know how to solve the recurrence in closed form, so 
we want to use a computer to grind out the values of $a_n$ for, 
say, $n \le 20$: $$a_1=1, \zj a_2 = 5, \zj a_3=75, \zj a_4= 6975, 
\zj a_5 = 48845925, 
 \zj \ldots.$$
(2)  We insist on knowing the answers exactly;
floating-point numbers of a fixed precision are not adequate.

The problem is that eventually the numbers will overflow the 
natural ``word size'' of the computer. 
 If integers are represented 
by 16 bits, the largest signed integer is $2^{15}-1$, the largest 
unsigned integer is $2^{16} -1$. If ``long'' (32-bit) integers are 
used, we can get up to $2^{32}-1$ unsigned. 

What happens when an integer variable overflows depends on the 
 programming language used.
Some systems will give an error message and abort the program.
In standard C, the storage of the number ``rolls over'' like a car's
odometer:   
 the most significant digits are lost, without warning. 
Programs like {\sl Maple\/} and {\sl Mathematica\/} do arithmetic 
with integers of (in principle) arbitrary length, 
 but let us assume that we are not using one of those 
 (else we would have no story to tell). 

The most obvious response to this situation would be to represent 
large integers as arrays of regular (say 16-bit) integers, and to 
program the arithmetical operations directly on them. 
 Each integer piece is treated exactly like a single digit in hand 
arithmetic. 
 For example, if $n = n_0 + 2^{16} n_1$ and $m = m_0 + 2^{16} m_1\,$,
then $nm = n_0 m_0 + 2^{16} (n_1 m_0 + n_0 m_1) + 2^{32} n_1 m_1\,$;
if we have $n_0\,$, etc., stored as 32-bit integers (that are actually
smaller than $2^{16}$), then $n_0 m_0$ is a 32-bit integer, and we can
break it up as $p_0 + 2^{16} p_1\,$, so that
$nm = p_0 + 2^{16} (p_0 + n_1 m_0 + n_0 m_1) + 2^{32} n_1 m_1\,$.
Next we can break up the coefficient of $2^{16}$ as $p_1 + 2^{16} p_2$
and throw $p_2$ into the coefficient of $2^{32}$.  And so on.
Addition and subtraction are easier, but similar; you need to handle
carries and borrows between adjacent 16-bit parts.

This method can be programmed straightforwardly enough,
but it has some disadvantages.
Multiplying two arrays of length $N$ requires $N^2$ multiplications 
of 16-bit numbers.  
Addition and subtraction require only $N$ operations, but these 
must be done sequentially, to handle the carries and borrows. 
 This is a disadvantage on a modern parallel computer.

In contrast, the modular method            
 to be described here requires only 
$N$ operations for input of size $N$, even for multiplication, and 
the operations are independent, so they can be performed in 
parallel. 
 (And, most importantly for our present purposes, the mathematical 
theory behind this method is much more interesting!) 

 \bigskip
 {\<\bf Modular (residue) arithmetic\quad}
 An integer $m>1$ defines a partition of $\z9Z$ into
equivalence classes
$$ [0],\ [1],\ \ldots,\ [m-1].$$
 (For more leisurely introductions to this subject, 
see the background reading 
recommended at the end of this article.)
Two numbers are ``equivalent (or congruent) modulo $m$'' 
if their difference is divisible by~$m$.
Thus $n \in [r]$ means that $n=qm +r$ with $0\le r < m-1$.
The modern computer notation is ``$r = n \mod m$'';
$r$ is called the ``residue'' of $n$ modulo~$m$;
 $m$ is called a ``modulus''\!.
(If $n_1$ and $n_2$ are in the same equivalence class, the
traditional mathematical notation is
``$ n_1 \zH n_2 \zT{ modulo } m$''\!,
without any implication that either $n_j$ lies in the range of the 
residues.) 

These classes can be added, subtracted, and multiplied
according to definitions
$$[a] + [b] = [a+b], \zj \zT{etc.}$$
We need to show that these operations are well-defined;
that is, the result of applying the definitions does not depend
on which elements $a$ and $b$ of the respective classes are chosen:

Suppose that $a = q_a m + r_a$ and $ b = q_b m + r_b\,$.
Then $a+b = (q_a +q_b) m + (r_a + r_b)$.
If $r_a + r_b \ge m$, we subtract $m$ from it to bring it back into
the proper range, and compensate by adding 1 to $q_a + q_b\,$.
If $r_a + r_b < m$, we do nothing.
In either case, we get the residue  of the sum as
$$r_{a+b} = (r_a + r_b) \mod m$$
regardless of what $q_a$  and $q_b$ were.

Similarly,
$$ ab = (q_a q_b m + r_a q_b + r_b q_a)m + (r_a r_b),$$
so $r_{ab} = (r_a r_b) \mod m$.

This arithmetical consistency makes it possible to blur the
distinctions among $a$, $[a]$, and $r_a\,$,
representing equivalence classes by residues and
understanding all arithmetic to be modulo~$m$.
 The equivalence classes are called ``the integers modulo~$m$'';
 the set of integers modulo $m$ is denoted by $\z9Z_m\,$.

The range 0 to $m-1$ for the residues
is the most obvious one to choose, but
actually any string of $m$ consecutive integers can be used
to represent the equivalence classes.
The range from $\lceil - m/2 \rceil$ to $\lfloor m/2 - \ze\rfloor$,
with 0 in the center, is often useful.
 One should really think of $\z9Z_m$ as forming a 
circle ($[m]$ being the same thing as $[0]$),
 and the labeling of its elements by a particular choice of 
residues as analogous to a choice of range $a \le \zy < a + 2\zp$ 
 for the angular coordinate on a circle. 

Observe that odometer-style integer overflow is actually an
implementation of arithmetic modulo $2^{16}$ 
(or $2^{32}$, as the case may be).
For unsigned integers, the range of the residues is 0 to
$2^{16} -1$;
for signed integers in ``two's-complement representation''\!,
the range is $- 2^{15}$ to $2^{15} -1$.
 (``One's-complement representation'' is a complication which
 we shall ignore.)

 \bigskip
 {\<\bf The Chinese remainder theorem\quad}
This theorem is called ``Chinese'' because a numerical example of it
is stated in a Chinese manuscript of circa 300 A.D.
(with author Sun Tsu), and the general case was stated and proved by
Ch'in Chiu-Shao in 1247 A.D.

Two integers are called ``relatively prime''
if their greatest common divisor is~1
(in other words, their prime factorizations have no number in common).

{\bf Chinese Remainder Theorem:}
Let $m_1\,$, \dots, $m_R$ be positive integers that are pairwise 
relatively prime:
$$ \gcd (m_j, m_k) = 1 \zj \zT{if } j \ne k.$$
Let $M = m_1 m_2 \cdots m_R\,$.
For any $R$-tuple of integers, $(u_1, \ldots, u_R)$,
there is exactly one integer $u$ such that
$$ 0 \le u < M \zj\zT{and}\zj u \zH u_j \zT{ modulo } m_j \zj
\zT{for each $j$}.$$

(In particular, if $u_j$ is restricted to the range 0 to $m_j-1$,
then $u_j = u\mod m_j\,$.
The same statement applies if the ranges of the residues are shifted,
or if the range of $u$ is shifted in a similar manner.)

 Before proving the theorem, let's look at an example.
 Suppose that $m_1 =12$ and $m_2=7$.
 Thus $M=84$.
 For any number $u$ between 0 and 83, say $u=49$, we can calculate 
the residues:
 $$\aligned u_1 = 1 \zj&\zT{because }49 = 4\zM 12 + 1, \\
            u_2 = 0 \zj&\zT{because }49 = 7\zM 7. \endaligned$$
 In fact, it is easy to see that all the numbers fit into a table,
 labeled by $u_1$ horizontally and $u_2$ vertically,
 starting with 0 in the upper left corner and moving diagonally 
downward, then ``wrapping'' to the top when the bottom is reached, 
and wrapping to the left whenever the right side is reached:
$$\matrix
 && C&C^\sharp&D&D^\sharp&E&F&F^\sharp&G&G^\sharp&A&A^\sharp&B \\
&&\text{\medbold 0}&\text{\medbold1}&\text{\medbold2}&\text{\medbold3}
&\text{\medbold4}&\text{\medbold5}&\text{\medbold6}&\text{\medbold7}
&\text{\medbold8}&\text{\medbold9}&\text{\medbold10}&\text{\medbold11} \\
\text{\tiny tonic} &\text{\medbold 0} & 0&49&14&63&28&77&42&7&56&21&70&35\\
\text{\tiny minor 2nd} &\text{\medbold 1} &36&1&50&15&64&29&78&43&8&57&22&71\\
\text{\tiny major 2nd} &\text{\medbold 2} &72&37&2&51&16&65&30&79&44&9&58&23\\
\text{\tiny minor 3rd} &\text{\medbold 3} &24&73&38&3&52&17&66&31&80&45&10&59\\
\text{\tiny major 3rd} &\text{\medbold 4} &60&25&74&39&4&53&18&67&32&81&46&11\\
\text{\tiny  4th} &\text{\medbold 5} &12&61&26&75&40&5&54&19&68&33&82&47\\
\text{\tiny minor 5th} &\text{\medbold 6} &48&13&62&27&76&41&6&55&20&69&34&83
\endmatrix$$
In other words, although we can continue to think of $\z9Z_{84}$
 as a circle, it also has the two-dimensional structure of 
a torus, the ``product'' of two circles of circumferences 12 and~7:
 $$\z9Z_{84} = \z9Z_{12} \zM \z9Z_7\,.$$

 This example has a musical interpretation.
Think of the integers modulo 84 as labeling musical notes, or the 
keys on a piano.
 (A standard piano has 88 keys, so there is an overlap of 4 keys at 
the ends.)
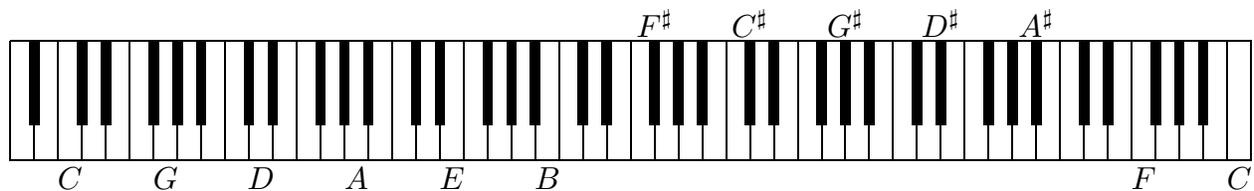
\newdimen\wkw\wkw=0.1249truein
\newdimen\wkh\wkh=5\wkw
\newdimen\bkw\bkw=0.4\wkw
$$\beginpicture
\setcoordinatesystem units <\wkw,\wkh>
 \setplotarea x from -2 to 50, y from 0 to 1
 \axis bottom ticks in length <\wkh> unlabeled from -2 to 50 by 1 /
 \axis top / 
 \linethickness=\bkw
 \putrule from -1 1 to -1 0.3
 \putrule from 1 1 to 1 0.3
 \putrule from 2 1 to 2 0.3
 \putrule from 4 1 to 4 0.3
 \putrule from 5 1 to 5 0.3
 \putrule from 6 1 to 6 0.3
 \putrule from 8 1 to 8 0.3
 \putrule from 9 1 to 9 0.3
 \putrule from 11 1 to 11 0.3
 \putrule from 12 1 to 12 0.3
 \putrule from 13 1 to 13 0.3
 \putrule from 15 1 to 15 0.3
 \putrule from 16 1 to 16 0.3
 \putrule from 18 1 to 18 0.3
 \putrule from 19 1 to 19 0.3
 \putrule from 20 1 to 20 0.3
 \putrule from 22 1 to 22 0.3
 \putrule from 23 1 to 23 0.3
 \putrule from 25 1 to 25 0.3
 \putrule from 26 1 to 26 0.3
 \putrule from 27 1 to 27 0.3
 \putrule from 29 1 to 29 0.3
 \putrule from 30 1 to 30 0.3
 \putrule from 32 1 to 32 0.3
 \putrule from 33 1 to 33 0.3
 \putrule from 34 1 to 34 0.3
 \putrule from 36 1 to 36 0.3
 \putrule from 37 1 to 37 0.3
 \putrule from 39 1 to 39 0.3
 \putrule from 40 1 to 40 0.3
 \putrule from 41 1 to 41 0.3 
 \putrule from 43 1 to 43 0.3 
 \putrule from 44 1 to 44 0.3
 \putrule from 46 1 to 46 0.3
 \putrule from 47 1 to 47 0.3 
 \putrule from 48 1 to 48 0.3
 \put{$C$} at 0.5 -0.15
 \put{$G$} at 4.5 -0.15
 \put{$D$} at 8.5 -0.15
 \put{$A$} at 12.5 -0.15
 \put{$E$} at 16.5 -0.15
 \put{$B$} at 20.5 -0.15
 \put{$F^\sharp$} at 25 1.15
 \put{$C^\sharp$} at 29 1.15
 \put{$G^\sharp$} at 33 1.15
 \put{$D^\sharp$} at 37 1.15
 \put{$A^\sharp$} at 41 1.15
 \put{$F$} at 45.5 -0.15
 \put{$C$} at 49.5 -0.15
 \endpicture$$
The ratio of the frequencies of the sounds produced by adjacent 
keys is $\root{12} \of 2$,
 so when we move up by 12 keys we reach the note of frequency 
twice the one we started from --- an interval of one 
``octave''\!. 
 If we started from a $C$, this note is also a~$C$.
 On the other hand, if we move up by 7 keys, we reach a note of 
frequency almost exactly $\frac32$ of the original one.
 This is called an interval of a ``fifth''\!.
 If we started from $C$ (as ``tonic''),
  this (so-called ``dominant'') note is  a~$G$.
 If we move up another 7 keys from $G$, we reach $D$;
 and continuing in this way we reach every one of the 12 notes of 
the scale exactly once before arriving back at a $C$.
 This structure is called ``the circle of fifths''\!.

 (You may wonder, since the basic numbers involved here are 12 
and~7, why the intervals are called ``octaves'' and ``fifths''\!.
 For complicated historical reasons, musicians have two strange 
rules for counting the notes between $C$ and $G$ or $C$:
 (1) don't count the black keys; 
 (2) count both endpoints (in mathematical terms, both [0] and $[m]$).
 The result is to convert the 12 to an 8 and the 7 to a~5.)

 It is important to understand that the existence of the circle of 
fifths depends crucially on the fact that 12 and 7 are relatively 
prime. 
  If we replaced the 7 by a 6, we would go from $C$ to 
$F^\sharp$ to $C$ again, without ever visiting the other notes.
 If we replaced 7 by 8 or 9 (not a divisor of 12 but not relatively 
prime to it either), the cycle would be longer but would still end 
prematurely.

 For an example of modular arithmetic in $\z9Z_{84}$, note that the 
residues of $v=2$ are
 $(v_1, v_2) = (2,2)$.
 As previously observed, the residues of 49 are $(u_1,u_2) = 
(1,0)$.
 Then
 $$(u_1,u_2) +(v_1, v_2) \zH ((u_1 + v_1)\mod 12 , (u_2+v_2)\mod 7)
  = (3, 2),$$
 and we see from the table that these are the residues of 51, which 
is indeed $49 + 2$.
 Similarly, 
$$(u_1,u_2) \zB (v_1, v_2) \zH (u_1  v_1\mod 12 , u_2v_2\mod7)
  = (2, 0),$$
which is the residue representation of 14;
 well, $49 \zB 2 = 98 = 84+ 14$, so $49 \zB 2$ does equal 14 modulo~84.

 For another example of the Chinese remainder theorem,
 see Example 14.18 in~[4], especially the table on 
p.~656, where the integers modulo $M=30$ are represented by their 
residues modulo 2, 3, and~5.

 A quick proof of the theorem is possible:
We can restate the conclusion this way:
For every list of residues, $(u_1, \ldots, u_R)$,
there is a unique $u\in \z9Z_M$ such that $u\mod m_j = u_j\,$.
 If $u$ and $u'$ are integers such that
$u \zH u' \zT{ modulo } m_j$ for $1\le j\le R$,
 then $u-u'$ is a multiple of each $m_j\,$.
 Therefore, $u-u'$ is a multiple of $M$.
 (This is the step where relative primality is used.)
This shows that the residues uniquely determine $u$ as a member
 of $\z9Z_M\,$.
 In other words, we have shown that the function mapping
 $u\in \z9Z_M$ into its list of residues, 
 $(u_1, \ldots, u_R)\in \z9Z_{m_1} \zM \cdots \zM \z9Z_{m_R}$,
 is injective.
 It remains to show that the function is surjective:
 every $R$-tuple of residues  corresponds to some $u\in \z9Z_M\,$.
But the number of $R$-tuples is $m_1 \cdots m_R = M$, the same as 
the number of integers modulo~$M$, so there is exactly one 
pigeon in every pigeonhole: the two sets are in one-to-one 
correspondence.

 As we saw earlier in the example,
 addition, subtraction, and multiplication modulo $M$ can be 
carried out by performing the corresponding operations on the 
residues modulo the respective $m_j\,$;
 for example,
 $$(u_1, \ldots, u_R) +  (v_1, \ldots, v_R)
 = ((u_1+v_1) \mod m_1, \ldots, (u_R+v_R) \mod m_R)$$
 yields the residue representation of $u+v$ modulo~$M$.
 We can think of the representation of an integer (less than $M$)
  by a string of residues as analogous to the representation of an 
  integer by a string of decimal digits, with the important 
  difference that under addition, subtraction, and multiplication
 the residues in each ``place'' combine only among themselves;
 there is no ``carry'' problem!

 \bigskip
 {\<\bf Implementation\quad}
 The practical point of this discussion should now be clear:
We choose $R$ relatively prime integers, $m_j\,$, 
 all fairly big but smaller than $2^{16}$,
 and we do computer arithmetic with arrays of $R$ integers,
 $(u_1, \ldots, u_R)$, modulo $(m_1, \ldots, m_R)$.
 These arrays represent all the integers from 0 up to 
one less than
 $M = m_1 \cdots m_R\,$, which may be much larger than $2^{32}$.
(In practice one may want to accommodate negative numbers by using 
the symmetric range,  
 $\lceil - M/2 \rceil \le u \le \lfloor M/2 - \ze\rfloor$, 
but for simplicity I shall not discuss the details of that.)
 This representation of large integers has the advantage previously 
advertised, that arithmetic can be performed in parallel on the 
residue components, without carries and borrows and without cross 
terms in multiplication.
 (It has some disadvantages, too, which will be mentioned later.)

 As a quick toy example, let's calculate $7!$ with respect to the 
moduli 13, 11, 9, and 7 --- without ever encountering a number 
larger than $256 = 2^8$.
 (This is possible because the moduli are all smaller than $2^4$.)
 The residues of 7 with respect to these moduli are $(7,7,7,0)$.
 For any smaller integer  $n$ the moduli are $(n,n,n,n)$.
 However, it is more efficient to combine some of them:
 $$ 2\zM 5 \zH (10, 10, 1, 3), \zJ 3\zM 4 \zH (12, 1, 3, 5).$$
Then
 $$ 2\zM3\zM4\zM5 \zH (120, 10, 3, 15) \zH (3, 10, 3, 1);$$
 $$ 6! \zH (18, 60, 18, 6) \zH (5, 5, 0, 6);$$
 $$7! \zH (35, 35, 0, 0) \zH (9, 2, 0, 0).$$
To check this, note that 
 $$\align 7!&=5040 \\
 &= 387\zM13 +9 \\
 &= 458\zM11 + 2 \\
 &= 560 \zM 9 +0 \\
 &= 6! \zM 7 +0. \endalign$$
 (Later I will explain how to find the 5040 from the $(9,2,0,0)$ 
without knowing the answer beforehand.)

 Several years ago an undergraduate student, Davin Potts, wrote 
under my direction some C{\tt++} functions to implement 
 residue arithmetic for large integers~[9].
 We recognized immediately that the problem was a natural one for 
object-oriented programming --- specifically for 
``polymorphism''\!, or ``operator overloading''\!.
The point is that in ordinary C code, when large integers are 
represented by arrays, one can no longer write operations like
 $${\tt c = a + b;}\,.$$
One needs to write something like 
 $${\tt c = sum(a,b);}\,,$$ 
 where {\tt sum} is the function (subroutine) one has written to 
carry out the addition.
 Even worse, if the integer is represented by a C structure that 
must be passed by reference, one might have
 $$\aligned &{\tt add(\&a, \&b, pc);} \\
            &{\tt c = {*}pc;} \,. \endaligned$$
In C{\tt++}, instead, we can define the $R$-tuples of residues 
to be a ``class'' named {\tt residue\_int}
 and can define ``+''\!, etc., to act on that class in the usual 
way:
$$\aligned&{\tt residue\_int\ a,\ b,\ c;} \\
          &{\tt c = a + b;}\,. \endaligned$$
 We also need to be able to add or multiply a {\tt residue\_int}
 by an ordinary integer;
 this is done by defining ``friend'' functions in C{\tt++}.

  The default moduli in Davin's program are the 5 largest prime 
 numbers less than $2^{16}$:
 $$ 65449, \zj 65479, \zj 65497, \zj 65519, \zj 65521.$$
 Their product is $1,204,964,463,846,332,731,259,513 \approx 10^{24}$,
  so we can represent all unsigned integers less than that value.
The program is flexible enough to accommodate various numbers of 
moduli.
 The residues themselves are represented as long integers (32 
bits), so that they can be multiplied without overflow.

 \bigskip
{\<\bf Reconstructing an integer from its residues\quad}
 Unfortunately, the list of residues of a large integer is not 
intuitively meaningful to a human reader.
 After a calculation one will probably want to convert from 
residues back into standard decimal notation.
 In the example of $\z9Z_{12} \zM \z9Z_7$ we could do that by 
reading the rectangular table, but that method is not practical if 
the moduli are large and there are more than 2 of them.
 
 A good reconstruction algorithm requires some more number theory.
I'll explain it in the context of our example, $m_1 =12$ and 
$m_2=7$.

 First we need to find a number $c$ that is the reciprocal 
(inverse) of one modulus modulo the other modulus. 
 I claim that 3 is the reciprocal of 12 modulo 7
 (in other words, $[12]^{-1} = [3]$ in the number system $\z9Z_7$).
 The proof is:
 $$ 12\zM3 = 36 = 5\zM 7 +1 \zH 1 \zT{ modulo } 7.$$
 (As an exercise, show that 7 is its own reciprocal modulo 12.
 The calculation is hidden in the previous text.
 Both reciprocals can also be read off from the table.
 Later I will mention a systematic way of finding reciprocals.)

 Knowing that  $c = [m_1]^{-1}$ in $\z9Z_{m_2}\,$,
 and given a residue list $(u_1,u_2)$, we can now define
 $$ v_2 = (u_2 - u_1) c \mod m_2 $$
 and get the corresponding integer modulo $M$ as
 $$ u = v_2 m_1 + u_1  \,.$$
 Proof:  Obviously $u\mod m_1 = u_1\,$.
 $$ \align u\mod m_2 &= [(u_2 - u_1) c m_1 \mod m_2] + [u_1 \mod 
m_2] \\
 &= [(u_2 -u_1) \mod m_2   ] + [u_1 \mod m_2] \\
 &= u_2 \mod m_2\,. \endalign$$
 Finally, one can check that $0\le u < M = m_1m_2\,.$

 Example:  In $\z9Z_{84}$ consider $(6, 4)$.
 Since $c =3$ in this case, $v_2 = -6 \mod 7 = 1$.
 Thus $u = 1\zM 12 + 6 = 18$.
 This agrees with our table ($F^\sharp$, major 3rd).

 This method can be iterated for more than two moduli.
 One needs first to find the reciprocal of each modulus with 
respect to each later modulus in the list:
 $$ c_{jk} m_j \zH 1 \zT{ modulo } m_k\,.$$
 Start the algorithm by setting 
 $(v_1, \ldots, v_R) =(u_1, \ldots, u_R)$.
 Then at step $j$ for $1\le j<R$, 
 $$\zT{replace $v_k$ by $(v_k-v_j) c_{jk} \mod m_k$}$$
  for $j<k\le R$.
 Finally, 
 $$u = v_1 + v_2 m_1 +v_3 m_2 m_1 + \cdots + v_R m_{R-1} \cdots 
m_1\,.  \eqno(*)$$
 I leave it as a big exercise to show in this way that $7! = 5040$.

 Of course, when this process is carried out in a real problem, $u$ 
and some of the terms in ($*$) must be expected to exceed the word 
size of the machine.
 It may be that at this point only the rough magnitude of $u$ is 
needed, and a conversion of the residue representation into a 
floating-point number, rather than an exact integer, will suffice.
 Davin wrote a C{\tt++} function to do that.

 This is a good place to mention the drawbacks of modular 
arithmetic.
 Although a modular representation is very convenient for addition, 
subtraction, and multiplication, it is not good at all for 
division, or for telling which of two integers is the larger, or 
for testing whether the integer is about to grow outside the 
allowed range set by the product of the moduli,~$M$.
 (These last two points are especially severe when the range 0 to 
$M-1$ is used, since it may be impossible to predict whether the 
result of a subtraction will be a negative number, which is not 
allowed.)
 To perform these operations one almost has to convert back to the 
form ($*$). 
(See, however, [7] and [8].)

 Another problem is that for a given underlying word size (say 
$2^{16}$), 
   there is an upper limit on the integers that can be represented, 
   since  there is a largest $M$ that can be obtained as a product 
   of prime  (or relatively prime) numbers less than the word size.
 This limitation is seldom important in practice.

 \bigskip
 {\<\bf The extended Euclidean algorithm\quad}
 One hole left in our description of the method is an algorithm for 
finding the reciprocals $c_{jk}\,$.
  An example of how this can be done appears in~[4] on 
  p.~649 (Example 14.13).
 The point is that the Euclidean algorithm for finding the greatest 
common divisor of $m_j$ and $m_k$ yields as  byproducts two 
integers $a$ and $b$ such that 
 $$am_1 + b m_2 = \gcd(m_1 ,m_2 ) \zj [{}= 1,\ \zT{since the $m_j$
 are relatively prime}].$$
It follows that $[a] = [m_1]^{-1}$ in $\z9Z_{m_2}$ and
 $[b] = [m_2]^{-1}$ in $\z9Z_{m_1}\,$.

 The ``extended Euclidean algorithm'' (which keeps track of the 
numbers $a$ and $b$ as well as the gcd)
 is extensively discussed by Knuth at the beginning of his first 
volume [5] as an example of the proof of validity of an algorithm by
 mathematical induction.
It is implemented in~[9].

 \bigskip\goodbreak
 {\<\bf Bibliographical remarks\quad}
 The primary research citation on the practical use of modular 
arithmetic for large numbers is~[1].
 The method is extensively discussed by Knuth in his second 
volume~[6];
 that book is my primary source for the Chinese remainder theorem 
and its proof.
 Szabo and Tanaka~[10] also studied computer applications of modular 
arithmetic, but they were more interested in working with small 
moduli and building the modular algorithms into the hardware.
Davis and Hersh~[3] give a readable introduction to the Chinese 
remainder theorem consisting primarily of interesting historical 
commentary.
 Grimaldi~[4] provides a standard textbook account of  
 residue arithmetic.

 \bigskip\goodbreak
 {\smalltype \<{\bf Acknowledgments.}
I thank Bob Blakley and Jack Borosh for introducing me to the 
residue method for handling large integers,
 and Davin Potts for my most successful student collaboration yet
(and for comments on the manuscript).
 \par}

 \bigskip\goodbreak
 \<{\smc references}
\medskip\frenchspacing\advance\parskip2\jot

 \item{1.} I. Borosh and A. S. Fraenkel, 
Exact solutions of linear equations with rational coefficients
by congruence techniques,
{\sl Math. Comput. \bf 20},
 107--112 (1966).

\item{2.} J. A. Campbell, Computation of a class of functions 
useful in the phase-integral approximation. I,
{\sl J. Comput. Phys. \bf10}, 308-315 (1972).


\item{3.}  P. J. Davis and R. Hersh, 
 {\sl The Mathematical Experience\/}
 (Birkh\"auser, Boston, 1981), pp. 187--195.

\item{4.}  R.  P. Grimaldi,
{\sl Discrete and Combinatorial Mathematics}, 
3rd ed.
(Addison--Wesley, Reading, 1994), Chap. 14.
 
\item{5.} D. E. Knuth, {\sl The Art of Computer Programming}, Vol.~1,
{\sl Fundamental Algorithms}, 2nd ed.  
 (Addison--Wesley, Reading, 1973),
pp.\ 14--17.

 \item{6.} D. E. Knuth, {\sl The Art of Computer Programming}, Vol.~2,
 {\sl Seminumerical Algorithms}, 2nd ed. (Addison--Wesley, Reading, 
1981), Sec. 4.3.2.

\item{7.} M.-L. Lin, E. Leiss, and B. McInnis,
Division and sign detection algorithms for residue number systems,
{\sl Comp. \& Maths. with Appls. \bf10}, 331-342 (1984).

\item{8.} M. Lu and J.-S. Chiang,
A novel division algorithm for the residue number system,
{\sl IEEE Transac. Comput. \bf41}, 1026--1032 (1992).

 \item{9.} D. M. Potts, 
 The RESIDUE\_INT software package,
 {\tt 
http://rainbow.uchicago.edu/\linebreak
 \~{}dmpotts/residue\_int/}~.
             
\item{10.} S. Szabo and R. Tanaka, 
 {\sl Residue Arithmetic and Its Applications to Computer 
Technology\/} (McGraw--Hill, New York, 1967).

\bye